\documentclass[preprint,aps,nofootinbib]{revtex4}
\usepackage{float}
\usepackage{graphicx}
\input{epsf.sty}
\usepackage{epsfig}

\begin{document}

\vskip 0.4cm
\title{ $R^2$ curvature-squared corrections on drag force }
\author{ Kazem Bitaghsir Fadafan\\Physics Department, Shahrood University of
Technology,\\ P.O.Box 3619995161 Shahrood, IRAN\\
bitaghsir@shahroodut.ac.ir}
\vspace*{2.0cm}
\begin{abstract}
The effect of  finite-coupling corrections to the drag force on a
moving heavy quark in the Super Yang-Mills plasma is investigated.
These corrections are related to curvature-squared corrections in
the corresponding gravity dual. The results are compared with the
dual gauge theory. It is shown that curvature-squared corrections
affect the drag force. It is shown that corrections to the drag
force depend on the velocity of the heavy quark. The diffusion
coefficient of non-relativistic heavy quarks is calculated from the
drag force. In addition, we also calculate the drag force on a
moving heavy quark in the Gauss-Bonnet background.
\end{abstract}

\vspace*{0.5cm}

 \maketitle
\section{Introduction}
The experiments of Relativistic Heavy Ion Collisions (RHIC),
collisions of gold nuclei at 200 GeV per nucleon have produced a
strongly-coupled quark-gluon plasma (QGP)\cite{Shuryak:2004cy}. The
AdS/CFT correspondence
\cite{Maldacena:1997re,Gubser:1998bc,Witten:1998qj, Witten:1998zw}
has yielded many important insights into the dynamics of
strongly-coupled gauge theories. It has been used to investigate the
hydrodynamical transport quantities in the various interesting
strongly-coupled gauge theories where the perturbation theory is not
applicable. Methods based on AdS/CFT relate the gravity in $AdS_5$
space to the conformal field theory on the 4-dimensional boundary.
It has been shown that an AdS space with a black brane is dual to
conformal field theory at finite temperature. Specially, dynamics of
open strings on a $AdS_5$ black brane background is related to the
quarks in the large N and large 't Hooft coupling limit of
4-dimensional the CFT dual theory at finite temperature. Then one
can calculate energy loss of quarks to the surrounding
strongly-coupled plasma.\\

One of the interesting properties of the strongly-coupled plasma at
RHIC is jet quenching of partons produced with high transverse
momentum. The jet quenching parameter controls the description of
relativistic partons and it is possible to employ the gauge/gravity
duality and determine this quantity at the finite temperature
theories. There has been the AdS/CFT calculation of jet quenching
parameter
\cite{Liu:2006ug,Buchel:2006bv,VazquezPoritz:2006ba,Caceres:2006as,
Lin:2006au,Avramis:2006ip,Armesto:2006zv,Argyres:2006vs,Argyres:2006yz,Argyres:2008eg}
and the drag coefficient which describes the energy loss for heavy
quarks in  $\mathcal{N}=4$ supersymmetric Yang-Mills theory
\cite{Nakano:2006js,Herzog:2006gh,Herzog:2006se,Caceres:2006dj,Matsuo:2006ws}.\\

 The universality of the ratio
of shear viscosity $\eta$ to entropy density $s$
\cite{Policastro:2001yc,Kovtun:2003wp,Buchel:2003tz,Kovtun:2004de}
for all gauge theories with Einstein gravity dual raised the
tantalizing prospect of a connection between string theory and RHIC.
The results were obtained for a class of gauge theories whose
holographic duals are dictated by classical Einstein gravity. But
string theory contains higher derivative corrections from stringy or
quantum effects, such corrections correspond to $1/\lambda$ and
$1/N$ corrections. In the case of $\mathcal{N}=4$ super Yang-Mills
theory, the gravity dual corresponds to type $\amalg B$ string
theory on $AdS_5\times S^5$ background. The leading order
corrections in $1/\lambda$ arises from the stringy corrections to
the low energy effective action of type $\amalg B$ supergravity,
$\alpha'^3 R^4$. Such corrections to the ratio of shear viscosity
$\eta$ to entropy density $s$ was calculated in
\cite{Benincasa:2005qc,Buchel:2004di}.\\

 Recently, the calculation
of $\frac{\eta}{s}$ for higher derivative gravity
 has been done by
\cite{Brigante:2007nu,Brigante008gz,Kats:2007mq}. In
\cite{Kats:2007mq}, they compute the effect of general $R^2$
corrections to the gravitational action in AdS space and show that
the conjecture low bound on the $\frac{\eta}{s}$ can be violated.
The computations of this ratio in an effective five-dimensional
setting have been discussed in \cite{Buchel:2008wy}. Regarding
\cite{Kats:2007mq} and motivated by the vastness of the string
landscape \cite{Douglas:2006es}, one can explore the modification of
drag force on a moving heavy quark in the strongly-coupled plasma.
We do not limit our study to specific known string theory
corrections, and consider the generic higher derivative terms in the
holographic gravity dual. In general, we do not know about forms of
higher derivative corrections in string theory, but it has been
known that by string landscape one expects that generic corrections
can occur.\\

In this paper, we investigate finite-coupling corrections to the
drag force on
 a moving heavy quark in the Super Yang-Mills plasma using
 AdS/CFT. These corrections are related to curvature-squared corrections in the corresponding gravity dual.
 It is shown that curvature-squared corrections affect the drag
force. However, we do not predict effect of them on the CFT dual
theory. because curvature-squared corrections are not the first
higher derivative corrections in type $\amalg B$ superstring theory.
The corrections to the drag force depend on the velocity of heavy
quark. It is shown in Fig. 1 that there is a critical velocity
($v_c$) such that for $v > v_c$ the corrections increase the drag
force. This phenomena might be important because at $v_c$
curvature-squared corrections have the minimum effects to the drag
force. The diffusion coefficient of non-relativistic heavy quark is
calculated from the drag force. In addition, we also
calculate the drag force on a  moving heavy quark in the Gauss-Bonnet background.\cite{Justin}.\\

While this paper was in the final stages of preparation, it came to
our attention that the drag force calculation for a Gauss-Bonnet
brane has simultaneously been done in \cite{Justin}.

\section{set up}
We study the gauge theory at finite temperature $T$ and assume the
geometry has a black hole.  In the gauge theory side, an external
quark can be introduced by a string that has a single end point at
the boundary and extends down to the horizon.  One can consider the
curvature-squared corrections on the AdS black brane solution of the
following 5-dim action
\begin{eqnarray}
S=\int d^5 x \sqrt{-g}\, [\frac{R}{2 \kappa}-\Lambda+c_1 R^2+c_2
R_{\mu\nu}R^{\mu\nu}+c_3
R_{\mu\nu\rho\sigma}R^{\mu\nu\rho\sigma}]\label{R2action},
\end{eqnarray}
where $c_i$ are arbitrary small coefficients and the negative
cosmological constant $\Lambda$ creates an AdS space with the radius
\begin{equation}
L^2=\frac{-6}{\kappa \Lambda}.
\end{equation}
For the well-known $\mathcal{N}=4$ SU(N) SYM, $c_3$ term does not
appear and one can use a field redefinition (as explained in
\cite{Kats:2007mq}),  but in general this does not need to be the
case and one can find a theory with $c_3>0$
\cite{Blau:1999vz,Fayyazuddin:1998fb}. It is important to point out
that the first higher derivative correction in weakly curved type
IIB backgrounds enters at order $R^4$, and not $R^2$, so we will not
predict effect of curvature-squared corrections on the
$\mathcal{N}=4$ SU(N) SYM. The black brane solution of $AdS_5$ space
with curvature-squared corrections is
\begin{eqnarray}
ds^2=-(\frac{r^2}{L^2})f(r)
dt^2+(\frac{r^2}{L^2})d\overrightarrow{x}^2+\frac{1}{(\frac{r^2}{L^2})f(r)}dr^2,
\end{eqnarray}
where
\begin{equation}
 f(r)=1-\frac{r_0^4}{r^4}+\alpha+\gamma \frac{r_0^8}{r^8}.
\end{equation}
and
\begin{equation}
\alpha=\frac{4 \kappa}{3 L^2} \left( 2 \left( 5 c_1 + c_2
\right)+c_3 \right),\,\,\, \gamma=\frac{4\kappa}{L^2} c_3.
\label{alpha}
\end{equation}
 and $r$ denotes the radial coordinate of the black brane geometry and
$t, \vec{x}$ label the directions along the boundary at the spatial
infinity. In these coordinates the event horizon is located at
$f(r_h)=0$ where $r_h$ can be found by solving this equation. The
boundary is located at infinity and the geometry will be as
asymptotically AdS with the radius L. The temperature is given by
\begin{equation}
T=\frac{r_0}{\pi L^2} \left( 1+ \frac{1}{4} \alpha - \frac{5}{4}
\gamma\right)\label{TR2}.
\end{equation}
The relevant string dynamics is captured by the Nambu-Goto action
\begin{eqnarray}
S=-T_0\int d\tau d\sigma\sqrt{-detg_{ab } },
\end{eqnarray}
and $X^\mu(\sigma, \tau)$ is a map from the string world-sheet into
the space-time. The coordinates $(\sigma, \tau)$ parameterize the
induced metric $g_{ab}$ on the string world-sheet. Defining $\dot X
=\partial_\tau X$, $X' =
\partial_\sigma X$, and $V \cdot W = V^\mu W^\nu G_{\mu\nu}$ where
$G_{\mu\nu}$ is the AdS black brane metric, we have
\begin{eqnarray}
-g=-detg_{ab }=(\dot X \cdot X')^2 - (X')^2(\dot X)^2.
\end{eqnarray}
One can make the static choice  $\sigma=r, \tau=t$ and following
\cite{Herzog:2006gh,Gubser:2006bz} focus on the dual configuration
of the external quark moving in the $x^1$ direction on the plasma.
The string in this case, trails behind its boundary endpoint as it
moves at constant speed $v$ in the $x^1$ direction
\begin{equation}
x^1(r,t)=v t+\xi(r),\,\,\,\,\,  x^2=0,\,\,\, x^3=0.
\end{equation}
 given this, one can find the lagrangian as the following
\begin{equation}
\mathcal{L}=\sqrt{-detg}=\sqrt{-G_{rr}G_{tt}-G_{rr}G_{xx}\,
v^2-G_{xx}G_{tt}\xi'^2},\label{generallag }
\end{equation}
The equation of motion for $\xi$  implies that $\frac{\partial
L}{\partial \xi'}$ is a constant. One can name this constant as
$\Pi_{\xi}$ and it is found as follows
\begin{equation}
\Pi_{\xi}=\xi' \frac{G_{tt}G_{xx}}{\sqrt{-g}},
\end{equation}
The drag force that is experienced by the heavy quark is calculated
by the current density for momentum along $x^1$ direction,
$\pi_{x}^{r}$, where it is found as the following
\begin{equation}
\pi_{x}^{r}=-T_0 \xi' \frac{G_{xx}G_{tt}}{\left(-detg_{ab}\right)},
\end{equation}
and drag force is obtained
\begin{equation}
\frac{dp_1}{dt}=\sqrt{-detg_{ab}}\, \pi_{x}^{r},
\end{equation}
As a result, the drag force is easily simplified  to
\begin{equation}
\frac{dp_1}{dt}=-T_0 \Pi_\xi\label{generaldrag}.
\end{equation}

\section{Drag force with the effect of curvature-squared corrections }
One can replace the components of AdS black brane metric and derive
the drag force. The lagrangian is calculated by (\ref{generallag })
and one can find
\begin{equation}
\mathcal{L}=\sqrt{1-\frac{v^2}{f(r)}+\frac{r^4}{L^4} f(r) \xi'^2},
\end{equation}
The constant of the motion is derived from equation of motion.
$\xi'$ can also be found easily by deriving this constant. The
result is as the following
\begin{equation}
\xi'^2=\frac{\Pi_\xi^2 (1-\frac{v^2}{f(r)})}{\frac{r^4}{L^4}f(r)
\left( \frac{r^4}{L^4}f(r)- \Pi_\xi^2 \right)},
\end{equation}
Near the horizon $r_h$, both numerator and  denominator are positive
for large $r$ and negative for small $r$. We are interested in a
string that stretches from the boundary to the horizon. In such a
string, $\xi'^2$ remains positive everywhere on the string.  As a
result both numerator and denominator change sign at the same point
and with this condition, one can find the constant of the motion.
The numerator changes sign at the following point
\begin{equation}
r_{critical}^4=\frac{r_0^4}{2 (1+\alpha -v^2)}+\frac{r_0^4 \sqrt{1-4
\gamma (1+\alpha -v^2)}}{2 (1+\alpha -v^2)}\label{rcrit},
\end{equation}
and it is clear that in the case of $\alpha=0=\gamma$, we will
notice the result of AdS black brane in
\cite{Herzog:2006gh,Gubser:2006bz}. The denominator changes sign at
the same point. As a result,
 the constant of the motion is derived as follows

\begin{equation}
\Pi_\xi^2=\frac{v^2 r_0^4}{2 L^4 (1+\alpha-v^2)}\left( 1+\sqrt{1-4
\gamma (1+\alpha-v^2)} \right).
\end{equation}

It is straightforward to find $\xi'$. The result is a lengthy
equation. The coefficients of $\alpha$ and $\gamma$ are small and
one can expand $\xi'$ to obtain the leading order terms

\begin{eqnarray}
\xi'&=&\frac{L^2 r_0^2 v}{r_0^4-r^4}+\alpha \left( \frac{L^2 r_0^2
v}{r_0^4-r^4} \right)\left(\frac{3 r^4-r_0^4}{2(r_0^4-r^4)}\right)
\nonumber
\\&+&\gamma \left( \frac{L^2 r_0^2 v}{r_0^4-r^4} \right) \left(\frac{r^4
r_0^4 (1-2v^2)+r_0^8(v^2-2)+r^8(v^2-1)}{2 r^8
(r^4-r_0^4)}\right)+\ldots\,.
\end{eqnarray}

The first term is the exact result in
\cite{Herzog:2006gh,Gubser:2006bz}. By integrating the above terms,
one can calculate $\xi$ , too.\\\\
Finally the drag force on a moving heavy quark through the
strongly-coupled with the effect of curvature-squared corrections is

\begin{equation}
\frac{dp_1}{dt}=-T_0 \left( \frac{v r_0^2}{\sqrt{2} L^2} \right)
\left(\frac{1+\sqrt{1-4\gamma
(1+\alpha-v^2)}}{(1+\alpha-v^2)}\right)^{1/2},
\end{equation}
We rewrite the drag force in terms of parameters of the
strongly-coupled plasma. The tension of the string is
$T_0=-\frac{1}{2 \pi \alpha'}$ and temperature of the plasma is
given by (\ref{TR2}). It is easy to find the parameters of drag
force in terms of temperature and coupling of the plasma. Using the
above result, the drag force in terms of the temperature and
coupling of the plasma is given by

\begin{equation}
\left(F_{drag}\right)_{R^2}=-\frac{\pi\,T^2\, \sqrt{g_{YM}^2
N}}{2}\left(\frac{v}{\sqrt{1-v^2}} \right)
\sqrt{\left(\frac{1+\sqrt{1-4\gamma (1+\alpha-v^2)}}{2\,\left(
1+\frac{\alpha}{4}-\frac{5\,\gamma}{4}
\right)^4\,\left(1+\frac{\alpha}{1-v^2}\right)}\right)}\label{drag}.
\end{equation}
We have considered the effect of curvature-squared corrections on
the AdS black brane metric in (\ref{drag}). This equation describes
the drag force on a moving heavy quark in the strongly-coupled
supersymmetric Yang-Mills plasma. The drag force on a moving heavy
quark in the dual gauge theory is derived from (\ref{drag}) if we
neglect the effect of curvature-squared corrections by plugging
$\alpha=\gamma=0$ in (\ref{drag}).
\begin{figure}
  \centerline{\includegraphics[width=3in]{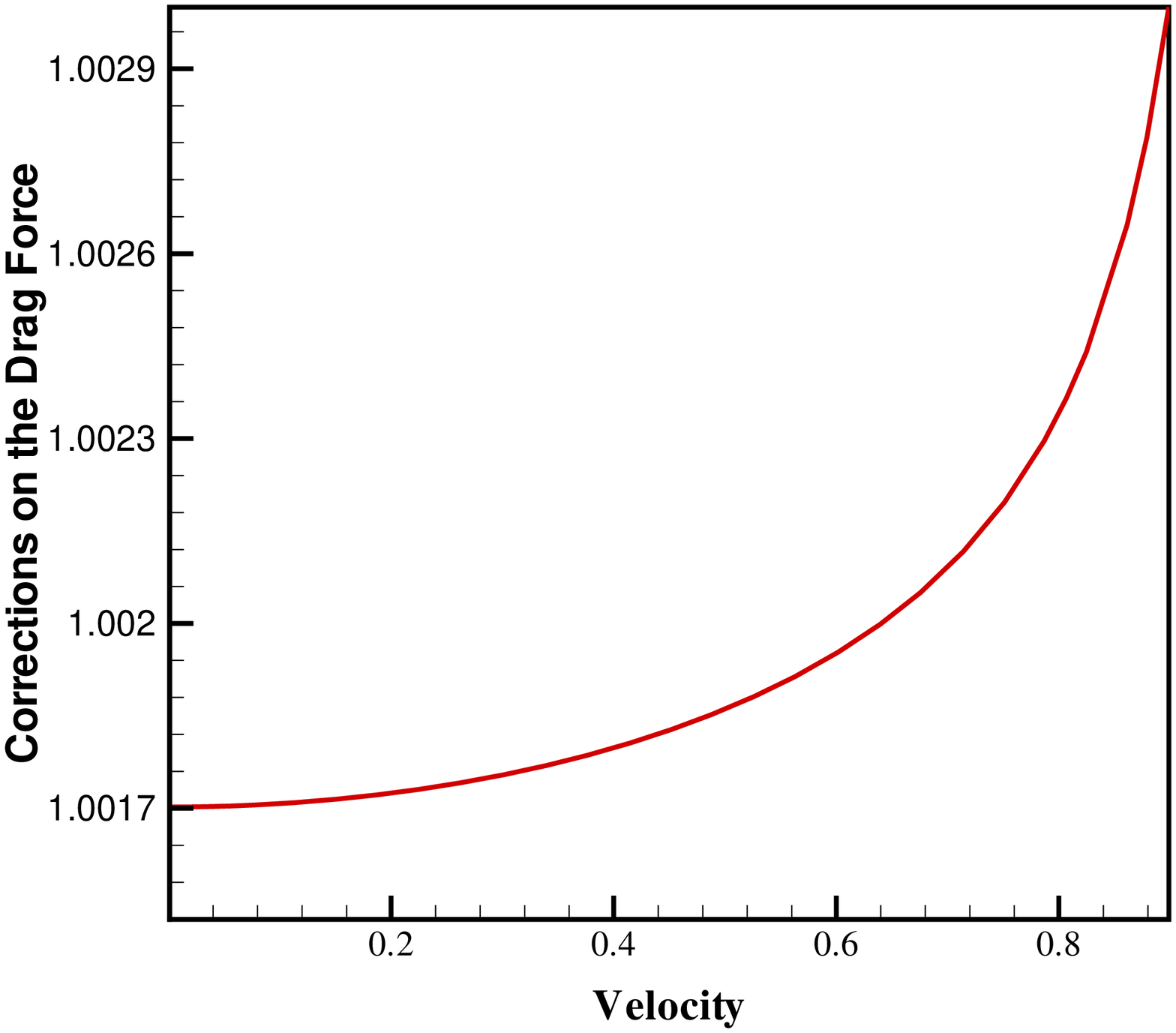} \includegraphics[width=3in]{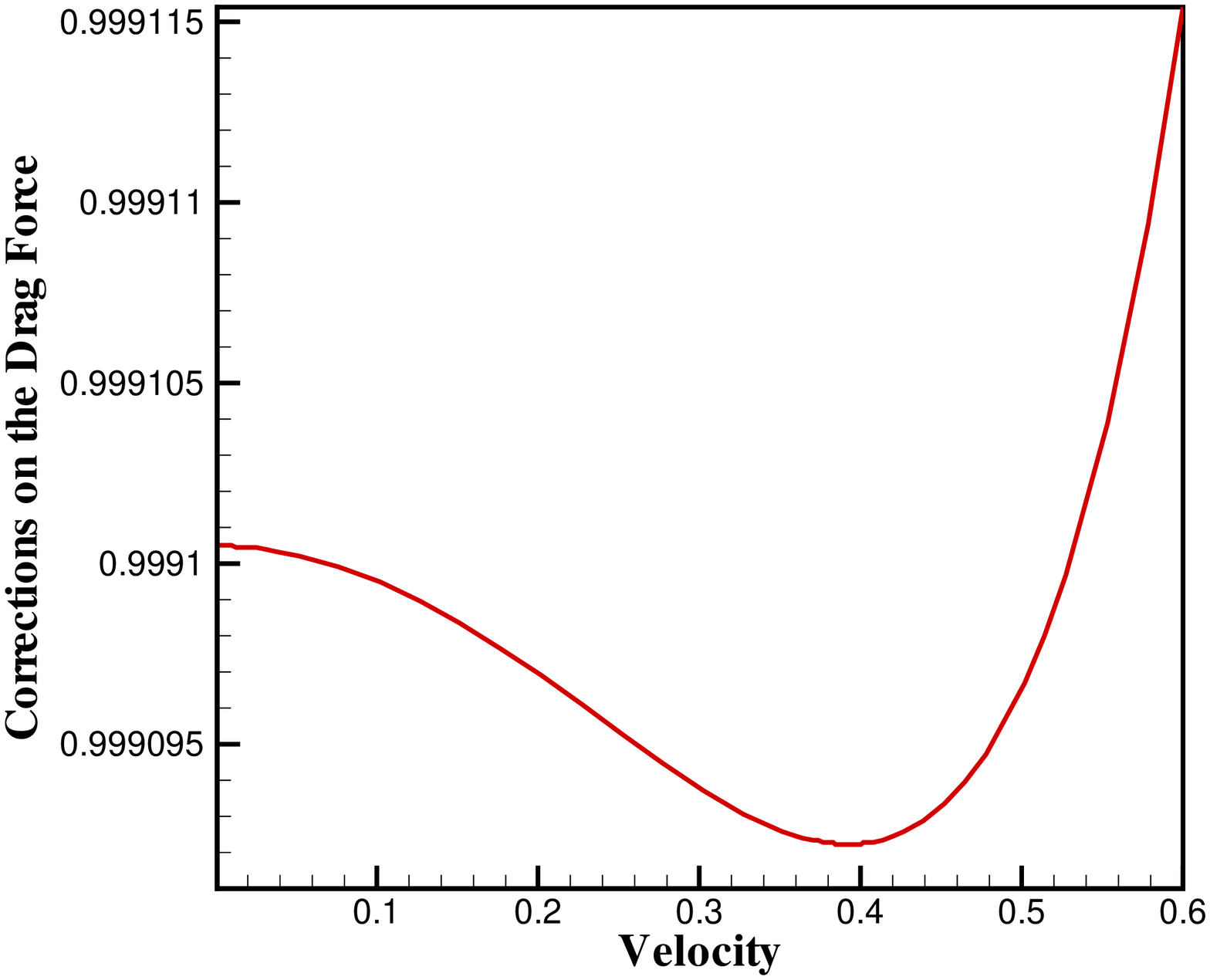}}
  \caption{The corrections to the drag force versus the velocity of the heavy quark with fixed small values of
  $\alpha$ and $\gamma$. Left:\,\,$\alpha=-0.0005$ and $\gamma=+0.0006$ so $T_{R^2}<
   T_{\mathcal{N}=4}$. Right:\,\,$\alpha=-0.0005$ and
$\gamma=-0.0007$ so $T_{R^2}>T_{\mathcal{N}=4}$.}\label{fig}
\end{figure}
 The effect of curvature-squared corrections in (\ref{drag}) appears in
the square-root.

We consider the square-root in (\ref{drag}) as the corrections to
the drag force. As we mentioned in (\ref{alpha}), parameters
$\alpha$ and $\gamma$ are small constants. One can discuss the
corrections to the drag force for different values of $\alpha$ and
$\gamma$. As an example, we have plotted in Fig. 1 the corrections
to the drag force versus the velocity of the heavy quark with fixed
small values of $\alpha$ and $\gamma$. In the left plot of Fig. 1,
$\alpha=-0.0005$ and $\gamma=+0.0006$ so $T_{R^2}<
T_{\mathcal{N}=4}$ where $T_{\mathcal{N}=4}$ is the temperature of
$\mathcal{N}=4$, SYM plasma and $T_{R^2}$ is the temperature of SYM
plasma with the effect of curvature-squared corrections,
respectively. As it is clear from the left plot, the corrections to
the drag force are increased monotonically with increasing the
velocity of the heavy quark and the corrections to the drag force
are larger than $\mathcal{N}=4$ case. By studying the derivative of
square-root in (\ref{drag}), one can see that it can be vanish. This
can be observed in the right plot of Fig.1. In this plot,
$\alpha=-0.0005$ and $\gamma=-0.0007$ so
$T_{R^2}>T_{\mathcal{N}=4}$. In this case, the corrections to the
drag force are smaller than $\mathcal{N}=4$ case. As it is obviously
seen from this plot, there is a critical velocity ($v_c$) such that
for $v>v_c$ the corrections increase the drag force. The critical
velocity and the condition on the parameters of $\alpha$ and
$\gamma$ can be found by studying the derivative of square-root in
(\ref{drag}). This phenomena might be important because at $v_c$ the
curvature-squared corrections have the minimum effects on the drag
force.

As a result, finite-coupling corrections affect the drag force on a
moving quark in the strongly-coupled plasma and depend on the
details of curvature-squared corrections in the corresponding
gravity dual. The drag force can be larger than or smaller than that
in the infinite-coupling case. We should emphasize that from our
results one cannot predict a result for $\mathcal{N}=4$ SYM because
 the first higher derivative correction in weakly curved type
IIB backgrounds enters at order $R^4$, and not $R^2$.
  \\
 \section{$R^2$ corrections on the diffusion coefficient of heavy quark in the SYM plasma }
The diffusion coefficient is a fundamental parameter of plasma at
RHIC for heavy quarks. It is known that the knowledge of the viscous
drag is equivalent to the knowledge of diffusion coefficient of
quark. The diffusion coefficient of non-relativistic heavy quarks
could be found from the drag force. We follow this approach in the
case of $\mathcal{N}=4$ Super
Yang-Mills theory. \\
The drag force on a heavy quark moving through the dual gauge theory
was calculated in \cite{Herzog:2006gh,Gubser:2006bz} as follows

\begin{equation}
\left(F_{drag} \right)_{\mathcal{N}=4}=\frac{\pi
\sqrt{g_{YM}^2N}}{2}\,T^2\,\frac{v}{\sqrt{1-v^2}}\,.\label{dragN=4}
\end{equation}
 The relaxation time can be derived from the above equation as the
 following

  \begin{equation}
t_{\mathcal{N}=4}=\frac{2\,m}{\pi\,T^2\, \sqrt{g_{YM}^2
N}}\label{time}.
\end{equation}
The diffusion coefficient is related to the temperature of the
plasma $T$, the heavy quark mass $m$ and the relaxation time $t_D$
as
 $D=\frac{T}{m}\,t_D$. It is straightforward to obtain the diffusion coefficient in the case of $\mathcal{N}=4$
 SYM plasma
\begin{equation}
D_{\mathcal{N}=4}=\frac{2}{\pi\,T\, \sqrt{g_{YM}^2 N}}\label{DN4}\,.
 \end{equation}
This result has been achieved with a different approach in
\cite{CasalderreySolana:2006rq}, for non-relativistic heavy quarks.
 Using the above approach, we can obtain the diffusion coefficient with the effect of
 finite-coupling corrections. If we neglect the
 squared-velocity terms of (\ref{drag}), the drag force for
 non-relativistic heavy quarks is given by
\begin{equation}
\left(F_{drag}\right)_{R^2}=-\frac{\pi\,p\,T^2\, \sqrt{g_{YM}^2
N}}{2\,m \sqrt{2}\,\left( 1+\frac{\alpha}{4}-\frac{5\,\gamma}{4}
\right)^2} \left(\frac{1+\sqrt{1-4\gamma
(1+\alpha)}}{(1+\alpha)}\right)^{1/2}\label{heavy-drag}.
\end{equation}
 The relaxation time of moving heavy quark can be calculated from
  the drag force. According to the above discussion,  one can obtain the diffusion coefficient with the effect of
finite-coupling corrections as follows
\begin{equation}
D=\frac{2 \sqrt{2}\,\left( 1+\frac{\alpha}{4}-\frac{5\,\gamma}{4}
\right)^2}{\pi \,T\, \sqrt{g_{YM}^2 N}}
\left(\frac{(1+\alpha)}{1+\sqrt{1-4\gamma (1+\alpha)}}\right)^{1/2}.
\end{equation} where $\alpha$ and $\gamma$ are small
constants. We can expand the above equation and keep the leading
order terms, the result would be
 \begin{equation}
D=\frac{2}{\pi\,T\, \sqrt{g_{YM}^2
N}}\left\{1+\alpha-2\gamma\right\}\label{leadingDR2}.
 \end{equation}
 We find that finite-coupling corrections affect the result of the dual gauge theory
 and one can argue about different signs of $\alpha$ and $\gamma$.
 Also the rate of change of the mean square transverse momentum of a
 non-relativistic heavy quark will be changed to
\begin{equation}
\frac{d}{dt}\,\,\langle (\vec{p}_\perp)^2
\rangle=\frac{4\,T^2}{D}\label{meanpath}.
\end{equation}
Jet quenching parameter, $\hat{q}$, as defined in
\cite{Baier:1996sk}, can be found by dividing (\ref{meanpath}) to
the velocity of the quark. Regarding the discussion of
 \cite{CasalderreySolana:2006rq} about the mass of heavy quark, the
 relaxation time $t_D=\frac{m}{T}D$ must be larger than the inverse
 temperature
 \begin{equation}
t_D \gg \frac{1}{T},
 \end{equation}
 which leads to
\begin{equation}
m \gg \frac{\pi\,T\, \sqrt{g_{YM}^2
N}}{2\,\left(1+\alpha-2\,\gamma\right)}.
 \end{equation}
One can tune mass by changing  $\alpha$ and $\gamma$.
\section{Gauss-Bonnet gravity background}
We study the black holes with higher derivative curvature in the AdS
space. In five dimensions, the most general theory of gravity with
quadratic powers of curvature is Einstein-Gauss-Bonnet (EGB) theory.
The exact solutions and thermodynamic properties of the black brane
in Gauss-Bonnet gravity were discussed in
\cite{Cai:2001dz,Nojiri:2001aj,Nojiri:2002qn}. Authors in
\cite{Brigante:2007nu,Brigante008gz} showed that for a class of CFTs
with Gauss-Bonnet gravity dual, the ratio of shear viscosity to
entropy density could violate the conjectured viscosity bound. The
computations of this ratio in an effective five-dimensional setting
have been discussed in \cite{Buchel:2008wy}. We try to understand
more about the drag force on a moving heavy quark in the boundary
gauge theory by string trailing in the Gauss-Bonnet gravity.

  The black brane solution in this geometry is given by
\begin{equation}
ds^2=-a \,\frac{r^2}{L^2}\, h(r)\, dt^2+\frac{dr^2}{\frac{r^2}{L^2}
h(r)}+\frac{r^2}{L^2} \,d\vec{x}^2\label{GBmetric},
\end{equation}
where
\begin{equation}
h(r)= \frac{1}{2\lambda_{GB}}\left[ 1-\sqrt{1-4 \lambda_{GB}\left(
1-\frac{r_+^4}{r^4} \right)}\right].
\end{equation}
In (\ref{GBmetric}), $a$ is an arbitrary constant which specifies
the speed of light of the boundary gauge theory and we choose it to
be unity. As a result at the boundary, where $r\rightarrow\infty$,
\begin{equation}
h(r)\rightarrow \frac{1}{a }, \,\,\,\,\, a= \frac{1}{2}\left(
1+\sqrt{1-4 \lambda_{GB}} \right)\label{a}.
\end{equation}
We assume $\lambda_{GB}\leq\frac{1}{4}$, the reason is that beyond
this point there is no vacuum AdS solution and one cannot have a
conformal field theory at the boundary. The curvature singularity
for $\lambda \geq 0$ occurs at $r=0$ and for $\lambda \leq 0$ the
curvature singularity is located at  $r=r_+\left(
1-\frac{1}{4\lambda_{GB} } \right)^{-1/2}$. The temperature is given
by
\begin{equation}
T=\frac{\sqrt{a}\,\, r_+}{\pi \,\,L^2}.
\end{equation}
Using (\ref{GBmetric}) the lagrangian would become
\begin{equation}
\mathcal{L}=\sqrt{a-\frac{v^2}{h(r)}+a \frac{r^4}{L^4} h(r) \xi'^2},
\end{equation}
and the constant of the motion is derived from equation of motion.
One may solve the relation for $\xi'$, the result is as the
following
\begin{equation}
\xi'^2=\frac{\Pi_\xi^2 (a-\frac{v^2}{h(r)})}{a \frac{r^4}{L^4}h(r)
\left( a \frac{r^4}{L^4}h(r)- \Pi_\xi^2 \right)}.
\end{equation}
As before, we look for the string that stretches from boundary to
horizon. Then numerator and denominator  change sign at the same
value
\begin{equation}
r_{critical}=\frac{\sqrt{a}r_+}{\left(a (a-v^2)+\lambda_{GB}
v^4\right)^{\frac{1}{4}}},
\end{equation}
with this condition, the constant of the motion is derived
 from denominator
 \begin{equation}
\Pi_\xi=\left( \frac{r_+^2 a}{L^2} \right) \frac{v}{\sqrt{a
(a-v^2)+\lambda_{GB} v^4}},\label{38}
\end{equation}
with this result, one can find the drag force from the above
equations. Following (\ref{generaldrag}), the drag force is derived
from (\ref{38}). The result is
\begin{equation}
\frac{dp_1}{dt}=-T_0 \left( \frac{r_+^2 a}{L^2} \right)
\frac{v}{\sqrt{a (a-v^2)+\lambda_{GB} v^4}}.
\end{equation}
We recall two useful formulas
\begin{equation}
L^4=g_{YM}^2 N  \alpha'^2, ~\ T= \frac{r_+ \sqrt{a}}{\pi L^2},
\end{equation}
where T is Hawking temperature or temperature of the plasma.
Plugging these relations into the drag force leads to the final
result for the drag force in the Gauss-Bonnet background
\begin{equation}
\left(F_{drag}\right)_{GB}=-\frac{\pi \sqrt{g_{YM}^2 N}}{2}T^2
\frac{v}{\sqrt{a(a-v^2)+\lambda_{GB}v^4}}\,.
\end{equation}

In this section, we considered the curvature-squared corrections and
found the drag force in the Gauss-Bonnet gravity. Now, we compare
this result with the drag force in AdS gravity where we know the
dual gauge theory exactly. The drag force on a moving heavy quark
through a thermal state of $SU(N)$ of $\mathcal{N}=4$ SYM theory is
given by (\ref{dragN=4}). One can compare the drag force in the
Gauss-Bonnet background, $\left(F_{drag}\right)_{GB}$ with the drag
force in the case of $\mathcal{N}=4$ SYM theory
,$\left(F_{drag}\right)_{\mathcal{N}=4}$ as the following
\begin{eqnarray}
\frac{\left(F_{drag}\right)_{GB}}{\left(F_{drag}\right)_{\mathcal{N}=4}}
&=&\frac{\,\sqrt{1-v^2}}{\sqrt{a(a-v^2)+\lambda_{GB}v^4}}
\nonumber\\&=&\frac{\sqrt{2}}{\sqrt{1-2
\lambda_{GB}(1+v^2)+\sqrt{1-4\lambda_{GB}}}}\label{fraction}\,.
\end{eqnarray}
where we have used the definition of $a$ in the equation
(\ref{a}).

It would be interesting to find values of $\lambda_{GB}$ where the
drag forces in two theories have the same value. It is clear from
the denominator of (\ref{fraction}) that at $\lambda_{GB}=0$, the
denominator is $\sqrt{2}$ and as a result the fraction of
(\ref{fraction}) will be unity! It is acceptable, because at this
value all curvature-squared terms will be zero. An important point
is that this value of $\lambda_{GB}$ is independent of velocity of
heavy quark and we will not expect a critical velocity. One can
obtain from the denominator of (\ref{fraction}) that if
$\lambda_{GB}>0$ the drag force on a moving heavy quark in the
Gauss-Bonnet gravity will be larger than $\mathcal{N}=4$ case. Also
for $\lambda_{GB}<0$, the drag force is smaller than $\mathcal{N}=4$
case.

 The above discussion for $\lambda_{GB}$ is
valid for the non-relativistic heavy quarks, where we neglect the
squared-velocity term in the (\ref{fraction}). The relaxation time
and diffusion coefficient of the non-relativistic heavy quark moving
through the strongly-coupled plasma in the Gauss-Bonnet gravity can
be obtained by following the calculations
of previous section. \\
\acknowledgments{ I would like to thank Urs. Achim Wiedemann who
directed me on QGP during my stay at CERN and many thanks from Luis
Alvarez-Gaume and  CERN theory group for hospitality. I would like
to thank S. Sheikh-Jabbari, H.Liu for very useful discussions and
specially thanks M. Edalati who introduced the work of Justin F.
$V\acute{a}$zquez-Poritz. I am grateful to Justin F.
$V\acute{a}$zquez-Poritz  for his kindly coordinate and would like
to thank H. Movahhedian and J.abouei for reading the manuscript.
This research was supported by shahrood university.}


\begin{thebibliography}{999}
\bibitem{Shuryak:2004cy}
  E.~V.~Shuryak,
  ``What RHIC experiments and theory tell us about properties of  quark-gluon
  plasma?,''
  Nucl.\ Phys.\  A {\bf 750} (2005) 64
  [arXiv:hep-ph/0405066].\\
  K.~Adcox {\it et al.}  [PHENIX Collaboration],
  ``Formation of dense partonic matter in relativistic nucleus nucleus
  collisions at RHIC: Experimental evaluation by the PHENIX  collaboration,''
  Nucl.\ Phys.\  A {\bf 757} (2005) 184
  [arXiv:nucl-ex/0410003].\\
  I.~Arsene {\it et al.}  [BRAHMS Collaboration],
  ``Quark gluon plasma and color glass condensate at RHIC? The perspective
  from the BRAHMS experiment,''
  Nucl.\ Phys.\  A {\bf 757} (2005) 1
  [arXiv:nucl-ex/0410020].\\
  J.~Adams {\it et al.}  [STAR Collaboration],
  ``Experimental and theoretical challenges in the search for the quark  gluon
  plasma: The STAR collaboration's critical assessment of the  evidence from
  RHIC collisions,''
  Nucl.\ Phys.\  A {\bf 757} (2005) 102
  [arXiv:nucl-ex/0501009].\\
\bibitem{Maldacena:1997re}
  J.~M.~Maldacena,
  ``The large N limit of superconformal field theories and supergravity,''
  Adv.\ Theor.\ Math.\ Phys.\  {\bf 2} (1998) 231
  [Int.\ J.\ Theor.\ Phys.\  {\bf 38} (1999) 1113]
  [arXiv:hep-th/9711200].
\bibitem{Gubser:1998bc}
  S.~S.~Gubser, I.~R.~Klebanov and A.~M.~Polyakov,
  ``Gauge theory correlators from non-critical string theory,''
  Phys.\ Lett.\  B {\bf 428} (1998) 105
  [arXiv:hep-th/9802109].
\bibitem{Witten:1998qj}
  E.~Witten,
  ``Anti-de Sitter space and holography,''
  Adv.\ Theor.\ Math.\ Phys.\  {\bf 2} (1998) 253
  [arXiv:hep-th/9802150].
\bibitem{Witten:1998zw}
  E.~Witten,
  ``Anti-de Sitter space, thermal phase transition, and confinement in  gauge
  theories,''
  Adv.\ Theor.\ Math.\ Phys.\  {\bf 2} (1998) 505
  [arXiv:hep-th/9803131].
\bibitem{Policastro:2001yc}
  G.~Policastro, D.~T.~Son and A.~O.~Starinets,
  ``The shear viscosity of strongly coupled N = 4 supersymmetric Yang-Mills
  plasma,''
  Phys.\ Rev.\ Lett.\  {\bf 87} (2001) 081601
  [arXiv:hep-th/0104066].
\bibitem{Kovtun:2003wp}
  P.~Kovtun, D.~T.~Son and A.~O.~Starinets,
  ``Holography and hydrodynamics: Diffusion on stretched horizons,''
  JHEP {\bf 0310} (2003) 064
  [arXiv:hep-th/0309213].
\bibitem{Buchel:2003tz}
  A.~Buchel and J.~T.~Liu,
  ``Universality of the shear viscosity in supergravity,''
  Phys.\ Rev.\ Lett.\  {\bf 93} (2004) 090602
  [arXiv:hep-th/0311175].
\bibitem{Kovtun:2004de}
  P.~Kovtun, D.~T.~Son and A.~O.~Starinets,
  ``Viscosity in strongly interacting quantum field theories from black hole
  physics,''
  Phys.\ Rev.\ Lett.\  {\bf 94} (2005) 111601
  [arXiv:hep-th/0405231].
\bibitem{Liu:2006ug}
  H.~Liu, K.~Rajagopal and U.~A.~Wiedemann,
  ``Calculating the jet quenching parameter from AdS/CFT,''
  Phys.\ Rev.\ Lett.\  {\bf 97} (2006) 182301
  [arXiv:hep-ph/0605178].
\bibitem{Buchel:2006bv}
  A.~Buchel,
  `On jet quenching parameters in strongly coupled non-conformal gauge
  theories,''
  Phys.\ Rev.\  D {\bf 74} (2006) 046006
  [arXiv:hep-th/0605178].
\bibitem{VazquezPoritz:2006ba}
  J.~F.~Vazquez-Poritz,
  ``Enhancing the jet quenching parameter from marginal deformations,''
  arXiv:hep-th/0605296.
\bibitem{Caceres:2006as}
  E.~Caceres and A.~Guijosa,
  ``On drag forces and jet quenching in strongly coupled plasmas,''
  JHEP {\bf 0612} (2006) 068
  [arXiv:hep-th/0606134].
\bibitem{Lin:2006au}
  F.~L.~Lin and T.~Matsuo,
  ``Jet quenching parameter in medium with chemical potential from AdS/CFT,''
  Phys.\ Lett.\  B {\bf 641} (2006) 45
  [arXiv:hep-th/0606136].
\bibitem{Avramis:2006ip}
  S.~D.~Avramis and K.~Sfetsos,
  ``Supergravity and the jet quenching parameter in the presence of R-charge
  densities,''
  JHEP {\bf 0701} (2007) 065
  [arXiv:hep-th/0606190].
\bibitem{Armesto:2006zv}
  N.~Armesto, J.~D.~Edelstein and J.~Mas,
  ``Jet quenching at finite 't Hooft coupling and chemical potential from
  AdS/CFT,''
  JHEP {\bf 0609} (2006) 039
  [arXiv:hep-ph/0606245].
\bibitem{Argyres:2006vs}
  P.~C.~Argyres, M.~Edalati and J.~F.~Vazquez-Poritz,
  ``No-drag string configurations for steadily moving quark-antiquark pairs in
  a thermal bath,''
  JHEP {\bf 0701} (2007) 105
  [arXiv:hep-th/0608118].
\bibitem{Argyres:2006yz}
  P.~C.~Argyres, M.~Edalati and J.~F.~Vazquez-Poritz,
  ``Spacelike strings and jet quenching from a Wilson loop,''
  JHEP {\bf 0704} (2007) 049
  [arXiv:hep-th/0612157].
\bibitem{Argyres:2008eg}
  P.~C.~Argyres, M.~Edalati and J.~F.~Vazquez-Poritz,
  ``Lightlike Wilson loops from AdS/CFT,''
  JHEP {\bf 0803} (2008) 071
  [arXiv:0801.4594 [hep-th]].

\bibitem{Nakano:2006js}
  E.~Nakano, S.~Teraguchi and W.~Y.~Wen,
  ``Drag Force, Jet Quenching, and AdS/QCD,''
  Phys.\ Rev.\  D {\bf 75} (2007) 085016
  [arXiv:hep-ph/0608274].
\bibitem{Herzog:2006se}
  C.~P.~Herzog,
  ``Energy loss of heavy quarks from asymptotically AdS geometries,''
  JHEP {\bf 0609} (2006) 032
  [arXiv:hep-th/0605191].
\bibitem{Caceres:2006dj}
  E.~Caceres and A.~Guijosa,
  `Drag force in charged N = 4 SYM plasma,''
  JHEP {\bf 0611} (2006) 077
  [arXiv:hep-th/0605235].
\bibitem{Matsuo:2006ws}
  T.~Matsuo, D.~Tomino and W.~Y.~Wen,
  ``Drag force in SYM plasma with B field from AdS/CFT,''
  JHEP {\bf 0610} (2006) 055
  [arXiv:hep-th/0607178].
\bibitem{Herzog:2006gh}
  C.~P.~Herzog, A.~Karch, P.~Kovtun, C.~Kozcaz and L.~G.~Yaffe,
  ``Energy loss of a heavy quark moving through N = 4 supersymmetric
  Yang-Mills plasma,''
  JHEP {\bf 0607} (2006) 013
  [arXiv:hep-th/0605158].
\bibitem{CasalderreySolana:2006rq}
  J.~Casalderrey-Solana and D.~Teaney,
  ``Heavy quark diffusion in strongly coupled N = 4 Yang Mills,''
  Phys.\ Rev.\  D {\bf 74} (2006) 085012
  [arXiv:hep-ph/0605199].
\bibitem{Benincasa:2005qc}
  P.~Benincasa and A.~Buchel,
  ``Transport properties of N = 4 supersymmetric Yang-Mills theory at  finite
  coupling,''
  JHEP {\bf 0601} (2006) 103
  [arXiv:hep-th/0510041].
\bibitem{Buchel:2004di}
  A.~Buchel, J.~T.~Liu and A.~O.~Starinets,
  ``Coupling constant dependence of the shear viscosity in N=4 supersymmetric
  Yang-Mills theory,''
  Nucl.\ Phys.\  B {\bf 707} (2005) 56
  [arXiv:hep-th/0406264].
\bibitem{Brigante:2007nu}
  M.~Brigante, H.~Liu, R.~C.~Myers, S.~Shenker and S.~Yaida,
  ``Viscosity Bound Violation in Higher Derivative Gravity,''
  arXiv:0712.0805 [hep-th].
\bibitem{Kats:2007mq}
  Y.~Kats and P.~Petrov,
  ``Effect of curvature squared corrections in AdS on the viscosity of the dual
  gauge theory,''
  arXiv:0712.0743 [hep-th].
\bibitem{Brigante008gz}
  M.~Brigante, H.~Liu, R.~C.~Myers, S.~Shenker and S.~Yaida,
  ``The Viscosity Bound and Causality Violation,''
  arXiv:0802.3318 [hep-th].

\bibitem{Buchel:2008wy}
  A.~Buchel,
  ``Shear viscosity of CFT plasma at finite coupling,''
  Phys.\ Lett.\  B {\bf 665} (2008) 298
  [arXiv:0804.3161 [hep-th]].


\bibitem{Douglas:2006es}
  M.~R.~Douglas and S.~Kachru,
  ``Flux compactification,''
  Rev.\ Mod.\ Phys.\  {\bf 79} (2007) 733
  [arXiv:hep-th/0610102].
\bibitem{Gubser:2006bz}
  S.~S.~Gubser,
  ``Drag force in AdS/CFT,''
  Phys.\ Rev.\  D {\bf 74} (2006) 126005
  [arXiv:hep-th/0605182].
\bibitem{Gubser:2006qh}
  S.~S.~Gubser,
  ``Comparing the drag force on heavy quarks in N = 4 super-Yang-Mills theory
  and QCD,''
  Phys.\ Rev.\  D {\bf 76} (2007) 126003
  [arXiv:hep-th/0611272].
\bibitem{Baier:1996sk}
  R.~Baier, Y.~L.~Dokshitzer, A.~H.~Mueller, S.~Peigne and D.~Schiff,
  ``Radiative energy loss and p(T)-broadening of high energy partons in
  nuclei,''
  Nucl.\ Phys.\  B {\bf 484} (1997) 265
  [arXiv:hep-ph/9608322].
\bibitem{Blau:1999vz}
  M.~Blau, K.~S.~Narain and E.~Gava,
  ``On subleading contributions to the AdS/CFT trace anomaly,''
  JHEP {\bf 9909} (1999) 018
  [arXiv:hep-th/9904179].
\bibitem{Fayyazuddin:1998fb}
  A.~Fayyazuddin and M.~Spalinski,
  ``Large N superconformal gauge theories and supergravity orientifolds,''
  Nucl.\ Phys.\  B {\bf 535} (1998) 219
  [arXiv:hep-th/9805096].
\bibitem{Cai:2001dz}
  R.~G.~Cai,
  ``Gauss-Bonnet black holes in AdS spaces,''
  Phys.\ Rev.\  D {\bf 65} (2002) 084014
  [arXiv:hep-th/0109133].
\bibitem{Nojiri:2001aj}
  S.~Nojiri and S.~D.~Odintsov,
  ``Anti-de Sitter black hole thermodynamics in higher derivative gravity  and
  new confining-deconfining phases in dual CFT,''
  Phys.\ Lett.\  B {\bf 521} (2001) 87
  [Erratum-ibid.\  B {\bf 542} (2002) 301]
  [arXiv:hep-th/0109122].
\bibitem{Nojiri:2002qn}
  S.~Nojiri and S.~D.~Odintsov,
  "(Anti-) de Sitter black holes in higher derivative gravity and dual
  conformal field theories,"
  Phys.\ Rev.\  D {\bf 66} (2002) 044012
  [arXiv:hep-th/0204112].

\bibitem{Justin}
  Justin F. V$\acute{a}$zquez-Poritz
  ``Drag force at finite t Hooft coupling from AdS/CFT,''
\end{thebibliography}
\end{document}